\documentclass[twoside]{ilcws08}
\usepackage[latin1]{inputenc}
\usepackage[dvips]{graphicx,epsfig,color}
\usepackage{wrapfig,rotating}
\usepackage{amssymb,amsmath,array}

\begin{document}
\title{
CALICE ScECAL Beam Test at Fermilab} 
\author{Satoru Uozumi {\it for the CALICE Collaboration}
\vspace{.3cm}\\
Physics Department, Faculty of Science, Kobe University \\
1-1 Rokkodai-cho, Kobe, Nada-ku, Hyogo 657-8501 Japan
}

\maketitle

\begin{abstract}

The scintillator-strip electromagnetic calorimeter (ScECAL) is one of the calorimeter technologies which can achieve fine granularity required for the particle flow algorithm.
Second prototype of the ScECAL has been built and tested with analog hadron calorimeter (AHCAL) and tail catcher (TCMT) in September 2008 at Fermilab meson test beam facility.
Data are taken  with 1 to 32~GeV of electron, pion and muon beams to evaluate all the necessary performances of the ScECAL, AHCAL and TCMT system.
This manuscript describes overview of the beam test and very preliminary results focusing on the ScECAL part.
\end{abstract}

\section{Introduction}
The scintillator-strip electromagnetic calorimeter (ScECAL) is one of the calorimeter technologies proposed in order to achieve necessary granularity for the particle flow algorithm \cite{PFA}.
Since there are many challenging issues to realize the ScECAL,
we have built a first small prototype and tested it with electron beams at DESY in 2007 spring \cite{ScECAL_DESYTB}.
After the first test, several studies and improvements has been done for the scintillator-strip and photo-sensor.
In order to establish the whole ScECAL technology including the improvements we have built a larger prototype and performed the beam test in September 2008 at Fermilab.
At the Fermilab beam test combined performance of the ScECAL with analog hadron calorimeter (AHCAL) and tail catcher (TCMT) is measured.
A first trial of $\pi^0$ reconstruction with the ScECAL is also examined.
In this paper overview of the ScECAL second prototype and some preliminary results of the Fermilab beam test are presented.

\section{The Scintillator-ECAL Second Prototype}
\begin{figure}[ht]
\begin{minipage}[t]{.6\textwidth}
\centerline{\includegraphics[width=0.99\columnwidth]{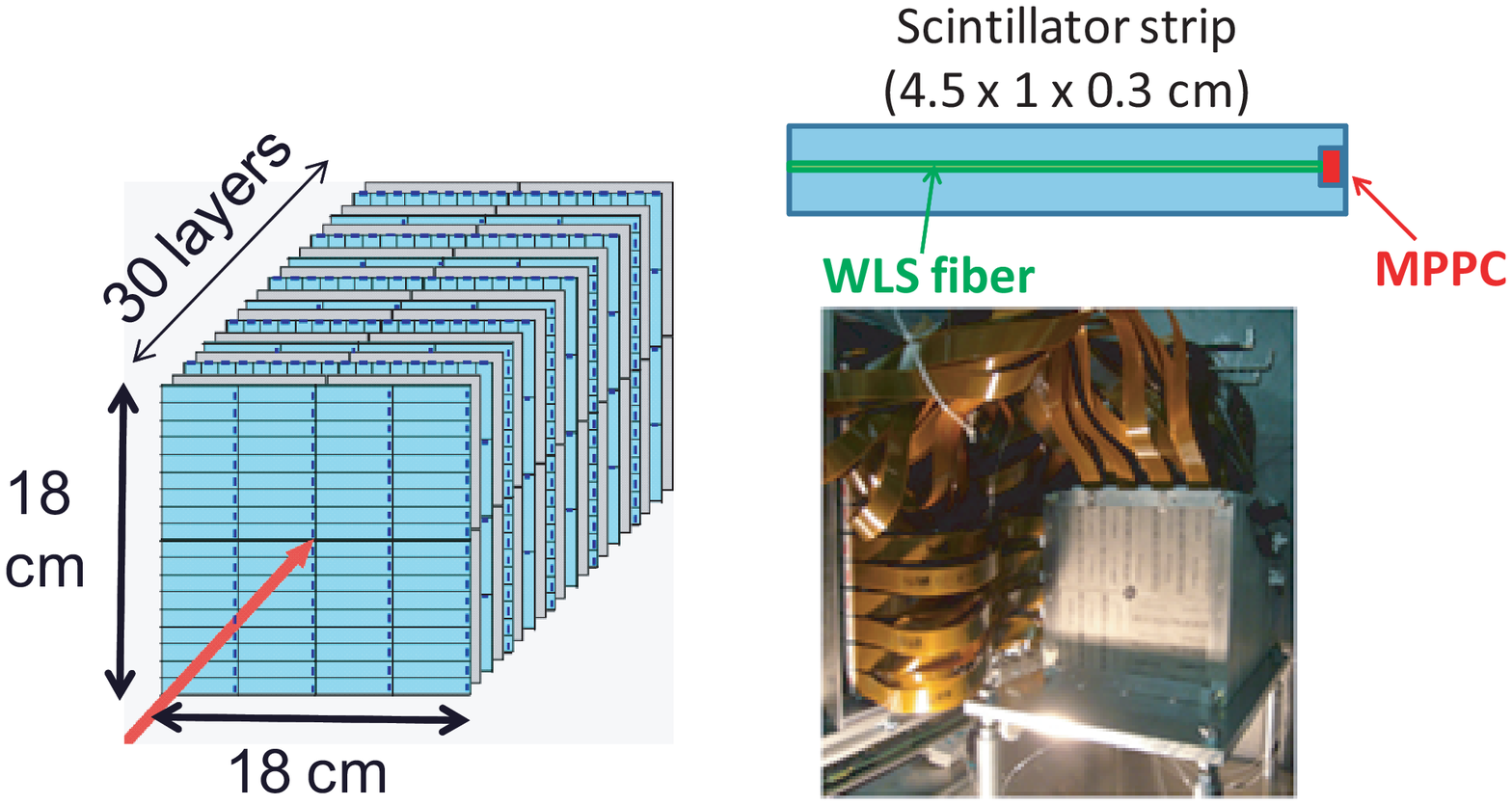}}
\caption{Structure and picture of the ScECAL second prototype.}\label{fig:prototype}
\end{minipage}
\hfill
\begin{minipage}[t]{.25\textwidth}
\centerline{\includegraphics[width=0.99\columnwidth]{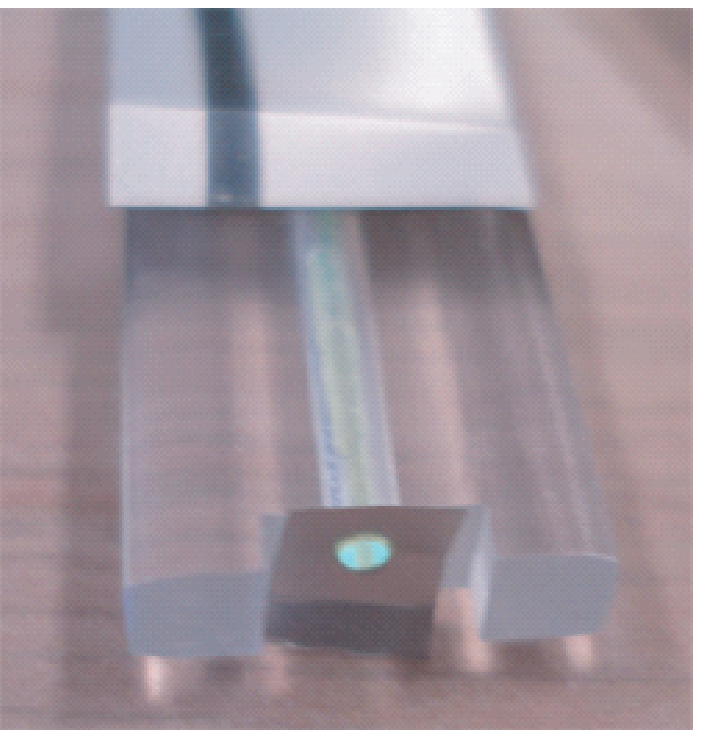}}
\caption{Reflector inserted between photo-sensor and scintillator to shield direct light from the scintillator.}\label{fig:directshield}
\end{minipage}
\end{figure}

Figure \ref{fig:prototype} shows structure of the ScECAL second prototype.
The prototype is made up of 30 pairs of absorber and scintillator layers.
Each scintillator layer consists of 72 scintillator strips made with extrusion method.
Size of each individual strip is $1\times4.5\times0.3$~cm.
At one edge of the strip a small semiconductor photo-sensor called MPPC \cite{MPPC} is attached.
Scintillation light signal caused by charged particles passing in the scintillator are absorbed into wavelength shifting fiber (WLS fiber) and guided into the photo-sensor.
As shown in Figure~\ref{fig:directshield}, a small reflector film with a hole fit to WLSF is inserted between the scintillator and photo-sensor.
This film shields photons which injects to the photo-sensor directly from the scintillator, not through the WLS fiber.
This shield effectively reduces undesired strip response non-uniformity.
In alternate layers scintillator strips are orthogonal to achieve 1~cm granularity both in two perpendicular (X and Y) directions.
The absorber layer is 3.5~mm thick and made from tungsten-cobalt alloy (W:Co=88:12 in weight ratio).
Overall size of the ScECAL is $18\times18\times20$~cm with total radiation length of 21$X_0$.

\section{CALICE Beam Test at Fermilab and Preliminary Results}

\begin{figure}[ht]
\centerline{\includegraphics[width=0.6\columnwidth]{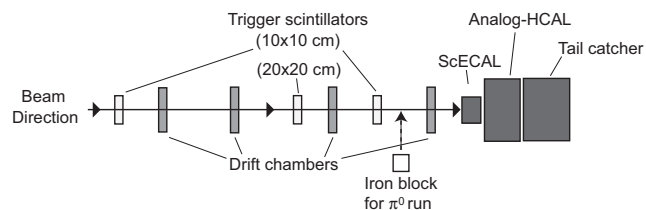}}
\caption{Schematic view of the beamline setup.}\label{fig:beamline}
\end{figure}

\begin{figure}[ht]
\begin{minipage}[t]{.3\textwidth}
\centerline{\includegraphics[width=0.99\columnwidth]{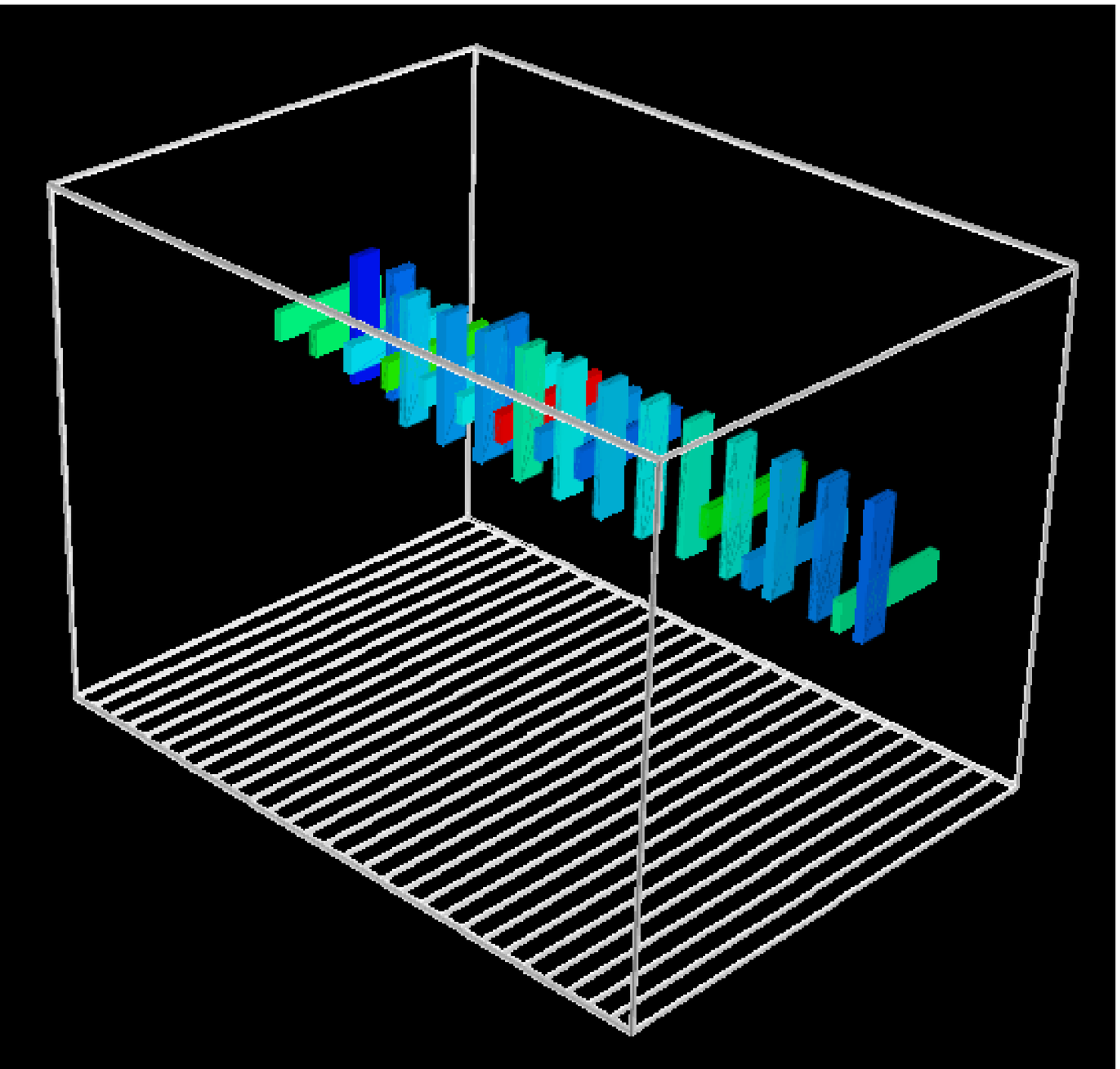}}
\end{minipage}
\hfill
\begin{minipage}[t]{.3\textwidth}
\centerline{\includegraphics[width=0.99\columnwidth]{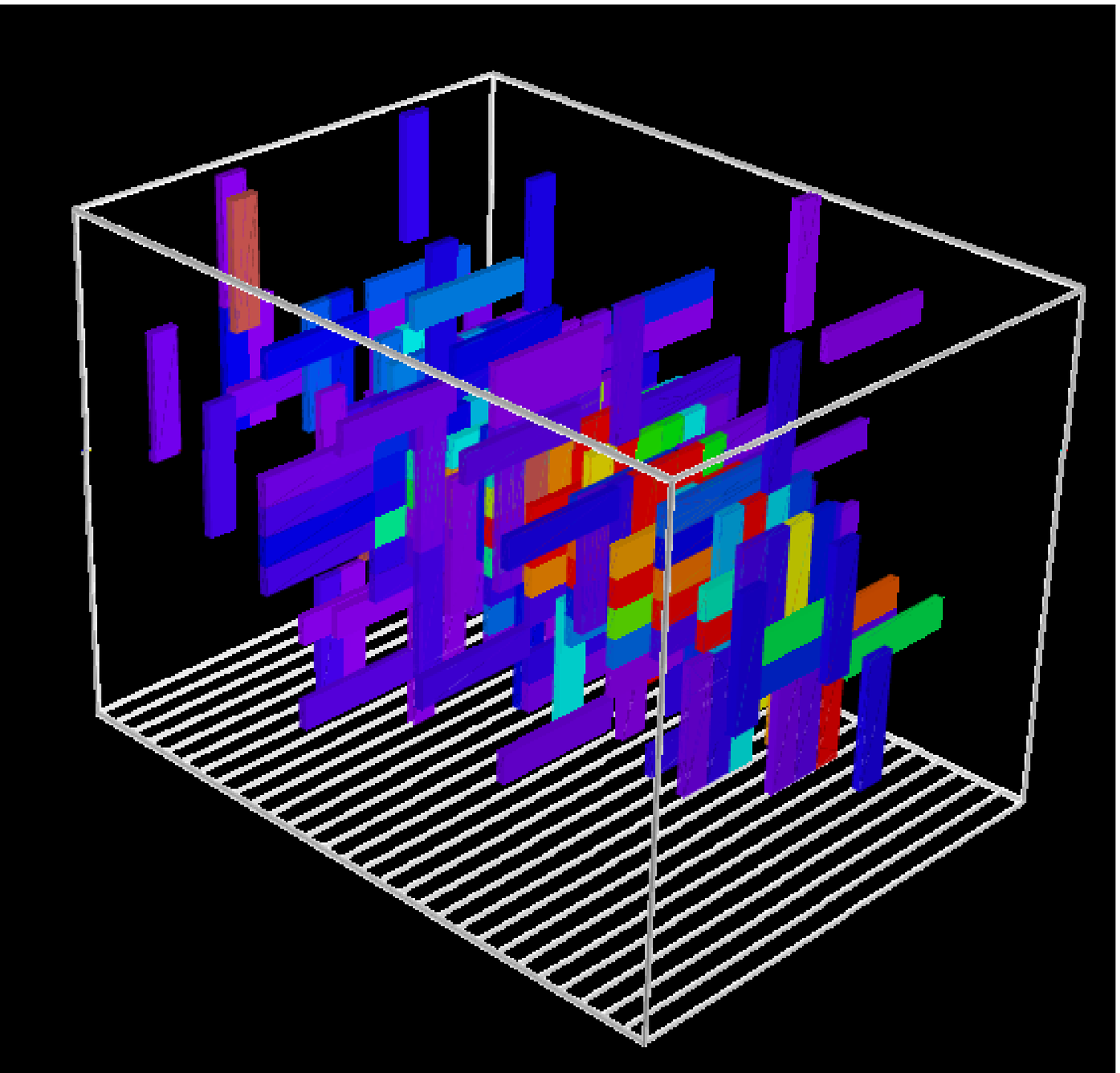}}
\end{minipage}
\hfill
\begin{minipage}[t]{.3\textwidth}
\centerline{\includegraphics[width=0.99\columnwidth]{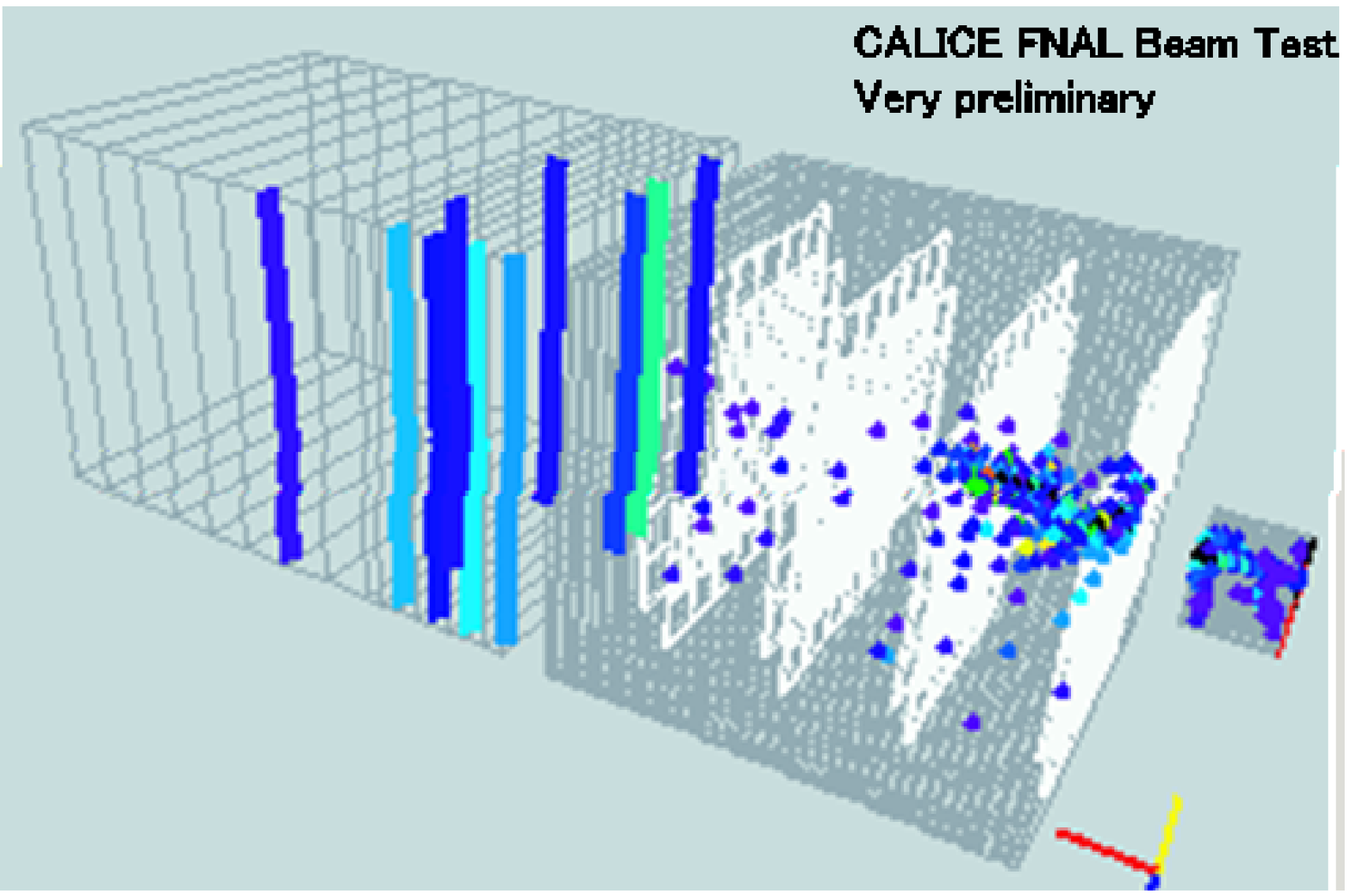}}
\end{minipage}
\caption{Event display of typical events. Left and center pictures are ScECAL display for a 32~GeV muon and 16~GeV electron events. Right one is ScECAL+AHCAL+TCMT combined view of a 16~GeV pion event.}\label{fig:evdisp}
\end{figure}

\begin{figure}[ht]
\centerline{\includegraphics[width=0.6\columnwidth]{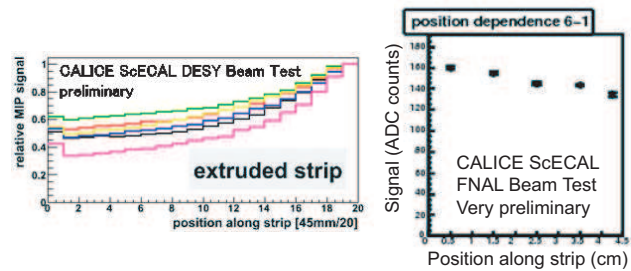}}
\caption{Scintillator-strip response to the minimum ionising particles as a function of hitting position.
Left plot is obtained at beam test of first prototype in 2007 at DESY.
With this plot the MPPC is attached at right end of the plot.
Right plot is a result from the 2008 Fermilab beam test.
The MPPC is located at position of zero (left end of the plot).
}\label{fig:nonuniformity}
\end{figure}

\begin{wrapfigure}{r}{0.45\columnwidth}
\centerline{\includegraphics[width=0.4\columnwidth]{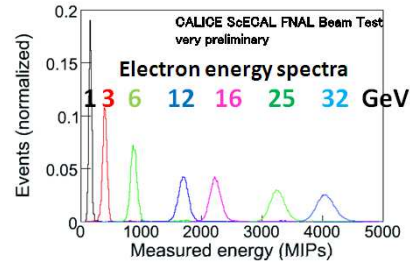}}
\caption{Electron energy spectra measured on the ScECAL. Unit of horizontal axis is given in MIP equivalent response.}\label{fig:spectra}
\end{wrapfigure}

The CALICE beam test has been carried out in September 2008 at Fermilab meson test beamline (MT6 section B).
The main goal of the test is to establish the ScECAL technology by evaluating performance of the ScECAL, AHCAL and TCMT combined system.
Description of the AHCAL and the TCMT can be found elsewhere \cite{AHCAL}.
Setup of the beamline is shown in Figure~\ref{fig:beamline}.
On upstream of the calorimeters there are 4 layers of drift chambers for beam position measurement.
Signal from 10$\times$10~cm or 20$\times$20~cm trigger scintillators are used for data taking trigger.
We have used electron, charged pion and muon beams with energy range of 1-32~GeV.
Event display of typical muon, electron and pion events are shown in Figure~\ref{fig:evdisp}.
After calibrating the response of all the scintillator-strips using muon beams, energy measurement of the electron beams with various energies has been done.
On Figure~\ref{fig:spectra} energy spectra of electrons measured by the ScECAL are displayed.
One can see clear Gaussian-shape spectra which proves feasibility of energy measurement by the ScECAL.
Analyses to evaluate energy resolution and response linearity are currently underway.
Figure~\ref{fig:nonuniformity} shows comparison of the position dependence between first and second prototype measured in 2007 DESY and 2008 Fermilab beam tests, respectively.
One can see that the ratio of scintillator-strip response with a MIP passing at farthest to closest side of the MPPC is improved from $\sim$40\% to $\sim$85\%.
Thanks to the treatment to shield the direct light from the scintillator not through the fiber (Figure~\ref{fig:prototype}), position dependence of the scintillator response has been significantly improved from the first prototype.

\begin{figure}[ht]
\centerline{\includegraphics[width=0.95\columnwidth]{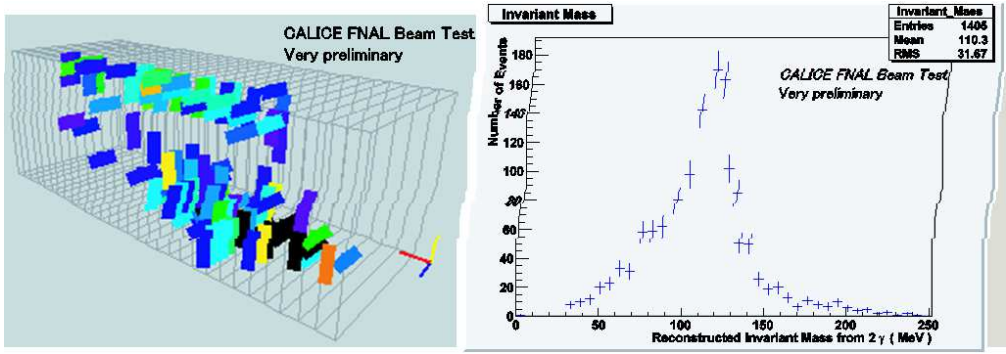}}
\caption{Event display of a typical $\pi^0$ candidate, and invariant mass distribution reconstructed from two electromagnetic clusters on the ScECAL.}\label{fig:pi0}
\end{figure}

We have also taken the $\pi^0$ data with 16-32~GeV $\pi^-$ beam.
As shown in Figure~\ref{fig:beamline}, during the $\pi^0$ runs an 10~cm thick iron block is inserted at 185~cm upstream of ScECAL (distance corresponds to ECAL inner radius of GLD-prime detector) to generate the $\pi^0$s with $\pi^-+\mbox{p/n}\rightarrow\pi^0+X$ reaction.
Pictures in Figure~\ref{fig:pi0} are an event display of a $\pi^0$ candidate and reconstructed mass distribution from two $\gamma$ clusters found on the ScECAL.

\clearpage

\section{Conclusion}
We have built the ScECAL second prototype and performed a beam test combining with AHCAL + TCMT system using $e^-$, $\pi^-$ and $\mu^-$ beams.
In energy range of 1-32~GeV we have observed clean electron energy spectra measured with the ScECAL.
The first attempt of $\pi^0$ reconstruction is also successful.
The test also gives us a lot of experience with the whole system including operation of the detector, electronics and various monitoring system.
Although analyses of the collected data are still extensively ongoing, first preliminary results already show enough performance and feasibility of the ScECAL.

\begin{footnotesize}

\end{footnotesize}


\end{document}